\documentclass[letterpaper]{article}
\usepackage{prepr}

\usepackage{amsmath,amsfonts} 
\usepackage{url}
\usepackage{booktabs} 
\usepackage{graphicx}
\usepackage{textcomp}
\usepackage{xcolor}
\usepackage{xspace}
\usepackage{hyperref}
\usepackage{balance}
\usepackage{colortbl}
\usepackage{tabularx}
\usepackage{tabulary}
\usepackage{booktabs}
\usepackage{multicol}
\usepackage{multirow}
\usepackage{rotating}
\usepackage{caption}
\usepackage{subcaption}
\usepackage[export]{adjustbox}
\usepackage{tabularx}
\usepackage{tikz}
\usepackage{float}
\usepackage[shortlabels]{enumitem}

\definecolor{commentGreen}{RGB}{128,127,41}
\definecolor{stringGreen}{RGB}{0,127,38}
\definecolor{keywordRed}{RGB}{151,0,11}
\definecolor{typePurple}{RGB}{128,3,82}
\definecolor{variableBlue}{RGB}{0,0,255}
\definecolor{annotationRed}{RGB}{198,4,38}
\definecolor{annotationBlue}{RGB}{32,21,223}

\definecolor{01gen}{RGB}{245,245,245}
\definecolor{02los}{RGB}{213,232,212}
\definecolor{03act}{RGB}{225,213,231}
\definecolor{04ref}{RGB}{218,232,252}
\definecolor{05misc}{RGB}{255,242,204}

\usepackage{tabularx}
\usepackage{booktabs}
\title{Bringing AI into the Classroom: A Structured Approach for Integrating  AI\\into Software Engineering Education}

\author{
\PREPauthor{Iris Groher}{Johannes Kepler University Linz, Institute of Business Informatics -- Software Engineering}{iris.groher@jku.at}
\PREPauthor{Michael Vierhauser}{University of Innsbruck, Department of Computer Science}{Michael.Vierhauser@uibk.ac.at}
\PREPauthor{Markus Weninger}{Johannes Kepler University Linz, Institute for System Software}{markus.weninger@jku.at}
}

\setvenue{\textbf{18th International Conference on Computer Supported Education}}
\setvenueone{the \textbf{18th Int'l Conf. on Computer Supported Education}}
\setyear{2026}
\setdoi{}

\newcommand{\repourl}{\url{https://github.com/TeachingAndLearningSciences/AIBlueprint}}
\newcommand{\marknum}[1]{\noindent\textbf{#1:}}

\begin{document}
\pagestyle{custom}
\maketitle

\begin{abstract}

The recent emergence of generative AI and Large Language Models (LLMs), particularly following the release of ChatGPT in late 2022, has significantly impacted both academic research and industrial practice.
This development has vast potential to impact educational practices, particularly computer science and software engineering courses.
Unfortunately, there is still a lack of actionable guidance on how to coherently integrate AI into computer science curricula.
In this paper, we therefore introduce the concept of \emph{AI-Blueprints}, a structured approach to integrating AI-related topics and activities into various computer science courses. We describe our approach and outline a structured process for creating new blueprints. 
Our vision is to provide these blueprints as open educational resources, allowing educators to adapt and integrate AI into diverse courses and topics. As a preliminary validation, we conducted semi-structured interviews with six university-level educators, collecting feedback on how our blueprints could help to integrate AI topics into existing courses. Based on this feedback, we outline plans for future research and expanding our AI-Blueprint concept.
\end{abstract}


\section{Introduction}
\label{sec:intro}

Artificial Intelligence (AI) has been an active research area for decades, largely driven by significant advances in machine learning algorithms and neural networks~\cite{sharifani2023machine}.
However, only recently has the rise of generative AI and transformer-based Large Language Models (LLMs) dramatically changed both academic research and industrial practice~\cite{kasneci2023chatgpt}.
With the release of OpenAI's ChatGPT in late 2022, LLMs have become omnipresent, making sophisticated language-processing capabilities and even programming assistance~\cite{sheese2024patterns} widely accessible. 

Given the rapid growth and broad accessibility of LLM-based tools, examining their implications for educational contexts has become increasingly important, particularly in software engineering (SE) education, which is currently under substantial transformation through the integration of LLMs~\cite{hou2024large-d64}.
Recent work in SE education has highlighted both the potential and limitations of LLMs, demonstrating their effectiveness in supporting code comprehension, automating formative feedback, and many other aspects~\cite{frankford24,guha2025future,kirova2024}. However, so far, integration of AI-related activities into existing SE courses and curricula is still sparse and unstructured at best, with actionable guidance for a coherent integration of LLM activities into computer science (CS) courses still missing.
Therefore, in this paper, we propose our initial ideas towards a novel concept of \emph{AI-Blueprints}.
Inspired by Use Cases from Requirements Engineering~\cite{cockburn1998basic}, which provide a structured template to describe a specific function of a system, we provide a structured approach to integrate AI-related topics and activities when planning CS courses.

Our contributions include a description of the AI-Blueprint format and how to apply it, complemented by a structured process for creating new blueprints~(\citesec{approach}), an initial report on our experience of applying the blueprint to three SE courses/topics, complemented by six semi-structured interviews with educators~(\citesec{validation}), and a discussion and a roadmap for future research~(\citesec{discussion}).

\section{Background \& Related Work}
\label{sec:background}

The ACM Curriculum guidelines~\cite{cs2023finalreport} represent a reference for structuring CS education. It has been widely recognized and adopted~\cite{skublewska2017acm}, and also serves as guidance for our envisioned blueprints.  The recent version CS2023\footnote{\url{https://csed.acm.org}} provides a modular framework of knowledge areas and units that define key topics, competencies, and learning outcomes, which can be adapted to local needs while maintaining global standards. Particular emphasis is placed on Software Engineering (SE) and Software Development Fundamentals (SDF), which cover essential competencies from programming basics and debugging to software design, testing, and project management. However, as noted by Vierhauser et al.~\cite{vierhauser2024towards}, few AI-related educational studies have targeted these areas. In this paper, we focus on SE and SDF, providing detailed lecture blueprints, and illustrate the adaptability of our proposed AI-Blueprint with an additional example from the Algorithmic Foundations (AL) area.

Different studies highlight the importance of integrating AI into SE curricula. Several works focus on \textit{student perceptions and usage}. Baresi et al.~\cite{ChatGPTEducation} examine LLM integration across universities, showing that students find them more useful for coding and debugging than for higher-level tasks, while Yabaku and Ouhbi~\cite{YabakuPerceptions} report frequent use for code optimisation and idea generation.

Other studies discuss \textit{course- and lecture-level integration strategies}. Abdelfattah et al.~\cite{AlyMethod} propose embedding ChatGPT in requirements engineering lectures to support user story and diagram creation, fostering active participation and critical thinking. Li et al.~\cite{LiStrategies} outline a framework aligned with the SE2014 Knowledge Areas, demonstrating integration into design and practical SE courses with activities such as code generation, bug detection, testing, and documentation.

Different authors address \textit{pedagogical frameworks and broader challenges}. Sah et al.~\cite{sah2024} compare dedicated AI/LLM courses, hybrid methods, and comprehensive programs, discussing their impact on technical and ethical competencies, as well as challenges such as limited resources, rapid technological change, and continuous training needs. Sengul et al.~\cite{SEEducationTrends} analyse conversational AI in SE, reporting early benefits for engagement in programming courses but also a gap between industry practice and higher education. Kirova et al.~\cite{kirova2024} call for adapting SE education to prepare graduates for an AI-driven future, including ethical awareness, while Dickey et al.~\cite{dickey2024} propose AI-Lab, a four-phase pedagogical framework that guides educators to integrate generative AI into programming courses.

Although all of these approaches provide important insights into how AI and LLM-specific approaches can be used in an educational context, a systematic framework that aligns existing course learning objectives with accompanying AI activities is still missing.
In the following, we introduce our AI-Blueprints and process as a structured response to this gap.

\section{AI-Blueprints: A Guiding Template}
\label{sec:approach}



Successful integration of AI into SE education requires a systematic methodology. Inspired by Use Case models in Requirements Engineering~\cite{cockburn1998basic}, we introduce AI-Blueprints, a structured template-based approach that supports educators in integrating AI-related content and activities into SE lectures. Each blueprint represents a reusable instructional pattern that combines existing learning objectives with AI concepts. The template structure enables adaptation across courses, educational levels, and institutions, and supports modular lecture design that allows incremental integration of AI activities without requiring a complete course redesign.


\subsection{AI-Blueprint Description}
\label{sec:blueprint-description}
As outlined in the simplified example blueprint\footnote{The full template, as well as all our example AI-Blueprints are available on GitHub~(\label{fnlabel}\repourl)} in \citetable{blueprint}, our proposed AI-Blueprint template provides a structured and systematic representation for educators to clearly outline, organise, and document the educational components augmented with AI. It comprises five main parts: (1) General Course and Lecture Description;  (2) 
Lecture Contents  \& Learning Objectives; (3) Lecture Activities \& Resources; (4) 
Lecture Reflection; and finally (5) Miscellaneous Info.
\vspace{0.5em}

\marknum{(1) General Course and Lecture Description} The first part provides general information about the course and the specific lecture. It aligns the lecture to one of the Knowledge Areas from the ACM CS2023 curriculum (e.g., Software Development Fundamentals -- SDF), supporting comparability across institutions, and clearly positions the educational context within established SE educational guidelines. It further refines this mapping to the specific Knowledge Unit within the selected area (e.g., Programming Basics). 
The Course Modalities specify the instructional format(s) (e.g., lecture, lab) and durations. Lecture Context situates the individual lecture within the broader course context by identifying its topic(s) and intended learning outcomes. Keywords list thematic terms for searching.
\vspace{0.5em}

\marknum{(2) Lecture Contents \& Learning Objectives} Beyond the basic content description of the first part, the second part facilitates the structured definition of learning objectives (LOs) tailored specifically for the dedicated course lecture. Our AI-Blueprints complement existing LOs by integrating additional AI-related objectives together with related tasks and activities without replacing established lecture content. 
It categorises both lecture-specific and AI-related LOs using Bloom's taxonomy and/or the \emph{skill levels} proposed in CS2023. Bloom’s (Revised) Taxonomy defines six hierarchical cognitive processes: Remember, Understand, Apply, Analyze, Evaluate, and Create, allowing educators to formulate measurable objectives at appropriate complexity levels~\cite{AndersonKrathwohl2001}. For the learning objectives, we use the skill levels  (Explain, Apply, Evaluate, Develop) from the CS2023 framework~\cite{cs2023finalreport}. We further complement these levels by common application types of AI in education~\cite{luo2025design} (Generate, Explain, Evaluate) for the AI-related learning objectives.
\vspace{0.5em}

\marknum{(3) Lecture Activities \& Resources} 
This part represents the core elements of our proposed AI-Blueprint. The activity-centered design emphasises the idea of ``active learning'' -- that students learn most effectively when they are directly engaged in meaningful, goal-oriented tasks~\cite{greer2019}. Describing small activities allows the adoption of individual parts of the blueprint, e.g., picking individual activities and integrating them into own lectures.
Each lecture activity is described following a standardised structure. To follow the principle of Constructive Alignment~\cite{biggs1996enhancing}, each activity is linked to the LOs and AI-LOs that this activity targets. This mapping ensures that the intended learning outcomes, the teaching/learning activities, the AI integration, and the assessment strategies are aligned. The \emph{Description} provides an overview of the student task in the form of a scenario, together with the intended collaborative format. Additionally, the \emph{Resources} part provides additional information about necessary teaching and learning materials, e.g., data, input, or material provided for the tasks, or specific sample prompts that can be used with an LLM. Finally, the \emph{Assessment} part provides a description and examples of how student performance can be evaluated, and how feedback should be provided.
In addition to in-class activities, the \emph{Assignment/Homework/Additional Activities} part can cover additional tasks targeted to be performed out of class, designed to reinforce both lecture-specific and AI-enhanced skills outside class. \vspace{0.5em}

\marknum{(4) Lecture Reflection} 
Complementing individual activities, it is essential for students to critically reflect on their learning experiences at the end of a lecture, with a focus on both their interaction with AI tools and the broader implications of AI integration in their work. This part guides structured student reflection, including both open-ended discussion questions, as well as questions fostering critical thinking, encouraging students to, on the one hand, share their experiences with the AI tools, and further ask students to evaluate trust, risk, and ethical implications.\vspace{0.5em}

\marknum{(5) Misc} This part lists any additional resources that support the lecture and reflection activities.

\subsection{AI-Blueprint Development Process}
\label{sec:blueprint-development}

To provide guidance and step-by-step instructions for developing new AI-Blueprints, a process for blueprint creation and adaptation for specific courses and lectures is necessary. Inspired by the AI-Lab framework~\cite{dickey2024}, we are convinced that detailed guidance can support educators in creating novel AI-Blueprints or in the modification of pre-existing ones to align with their particular course requirements.
As part of this, we have drafted an initial outline, sketching out the core steps necessary for using our proposed AI-Blueprint template. 
\vspace{0.5em}

\noindent\textbf{(1) Select Course and Lecture}: 
To identify courses and lectures, educators should first identify LOs in their courses that require skills that could be deepened through AI-assisted tasks. Educators might also think of lectures and topics in which students struggle with comprehension or productivity. In addition, educators can search our online repository\footref{fnlabel} to find AI use cases in SE teaching and adapt them to their context.
\vspace{0.5em}

\marknum{(2) LO- and AI-LO-Definition}
To integrate AI-related activities into a chosen lecture, we follow a learning-objective-centered approach. First, \emph{LOs should be assessed and selected}. LOs are widely recognised as a key instrument for defining structured and clearly articulated learning outcomes, and have become a standard component in many curriculum development processes~\cite{starr2008bloom,herring2000role}. 
Once appropriate LOs have been identified, corresponding \emph{AI-related LOs can be defined}. These supplementary LOs explicitly target AI skills and knowledge, and may define new LOs, or serve as extensions of existing course objectives. \vspace{0.5em}

\marknum{(3) Prepare (AI) Activities} Once LOs and AI-LOs are identified and defined, the blueprint can be created and populated with \emph{specific tasks and activities where AI tools and approaches can be used}.
Additionally, define assessment criteria by specifying both formative and summative assessments, including diagnostic questions and identification of potential misconceptions. Employ constructive alignment to ensure that all defined LOs, both course-specific and AI-related, are appropriately targeted and assessed in at least one activity.
Once activities and corresponding assessment criteria are defined, \emph{appropriate AI tools and technologies can be selected} based on their alignment with defined AI tasks and suitability for the student's skill level. Prepare these tools by configuring them for instructional use, by developing user guides, tutorials, or demonstrations.
 \\[-1em]
 
\textit{Lecture Material Preparation:}
Once concrete activities and tools are selected, \emph{corresponding lecture materials can be developed}. 
Materials should go beyond general instructional design by explicitly addressing teaching with and about AI. Prepare lecture slides, prompts, and exercises that teach students core AI concepts such as prompt engineering, LLM capabilities, and limitations of AI-generated outputs. Include examples that demonstrate how AI tools can support SE tasks, while also highlighting failure cases or misleading outputs.
\vspace{0.5em}

\marknum{(4) Post Activity Preparation}
Besides the AI-related tasks and activities, it is important that students learn how to \emph{critically reflect and discuss} the usage of AI tools. While, for example, LLMs can be valuable tools in educational and SE contexts, they are not infallible.
Therefore, it is important to not only teach students how to use these tools and create prompts but to foster discussion on how to interpret results, assess their quality, and improve in case they are not satisfactory. Post-lecture activities should facilitate student reflection and critical thinking.  
\vspace{0.5em}

\marknum{(5) Assess and Update Materials}
Finally, once a blueprint is created and used in practice, the experiences and feedback should be used to update and continuously improve the lecture tasks. Evaluate the effectiveness of the implemented AI tasks and update materials accordingly, incorporating feedback derived from formative and summative assessments to ensure continuous improvement and alignment with LOs.

\begin{table*}[h!]
\centering
\scriptsize
\begin{tabularx}{\textwidth}{|p{2.8cm}|X|}
\hline

\multicolumn{2}{|c|}{\cellcolor{01gen} \textbf{1. General Course and Lecture Description}} \\\hline

  ACM Category & Software Development Fundamentals (SDF) \\ \cline{1-2}
  Knowledge Unit & SDF: Programming Basics \\ \cline{1-2}
  Course Overview & The introductory Python for Business Administration course offers students a foundation in core programming concepts—such as variables, control flow, and functions—based on examples from a business context. It emphasizes practical problem-solving skills and data-driven decision making within a business context. \\ \cline{1-2}
  Course Modalities & Lab 90 minutes \\ \cline{1-2}
  Lecture Context & The pandas lecture introduces students to the pandas library's key data structures (Series and DataFrame) and fundamental operations—such as filtering, grouping, basic aggregation, plotting—enabling them to efficiently manipulate and analyze tabular data relevant to business scenarios. \\ \cline{1-2}
  Keywords & Python, Pandas, Data Analytics
\\ \hline

\multicolumn{2}{|c|}{\cellcolor{02los} \textbf{2. Lecture Contents \& Learning Objectives}} \\\hline

 Lecture-related LOs &
Students are able to ... \newline
-- LO\_C1: Describe structure/components of a DataFrame, and explain key operations. [UNDERSTAND]\newline
-- LO\_C2: Load tabular data and clean it. [APPLY] 
 \\ \cline{1-2}
 AI-related LOs & 
-- LO\_AI1: Use AI to explain pandas structures (index, columns, dtypes). [UNDERSTAND] [EXPLAIN]\newline
-- LO\_AI2: Use AI to generate/modify pandas code (load, clean, group, visualize). [APPLY] [GENERATE]\newline
-- LO\_AI3: Critically assess AI-generated workflows and refine. [EVALUATE] 
\\ \hline

\multicolumn{2}{|c|}{\cellcolor{03act} \textbf{3. Lecture Activities \& Resources }} \\\hline

Description & 
Students use AI to explore, generate, and refine pandas code. They explain structures, execute AI suggestions, and co-create merged code from AI and hand-written examples.\\ \cline{1-2}

 Activities & 

  \begin{tabular}[t]{
    | p{1.5cm}| p{9.8cm}|}
  \firsthline
  \multicolumn{1}{|l|}{\cellcolor{03act}\bfseries 3.1-Activity } 
    & \multicolumn{1}{l|}{\cellcolor{03act}\bfseries AI-based Explanation of Pandas Concepts}\\ 
  \hline
  Co. Algmt. & LO\_AI1  \\
  \hline
  Description & Use an AI assistant to explore DataFrame structure and operation semantics. Individually, students write a natural-language prompt such as P1. In pairs, they submit their prompt to ChatGPT (or Github Copilot) and read the explanation.  \\
    \hline
  Resources & P1: \emph{“You are a pandas expert and instructor, known for clear and concise explanations of DataFrame internals and operations. Explain what df.dtypes tells me about my DataFrame, and how I might change a column’s dtype.”} [EXPLAIN] [Persona Pattern] \\
    \hline
  Assessment & Each pair quickly presents something they learned that was new or well explained and something they found confusing, wrong, or incomplete.  \\
  \hline
  \end{tabular}
  \begin{tabular}[t]{
    | p{1.5cm}| p{9.8cm}|}
  \firsthline
  \multicolumn{1}{|l|}{\cellcolor{03act}\bfseries 3.2-Activity } 
    & \multicolumn{1}{l|}{\cellcolor{03act}\bfseries AI-based Code Generation}\\ 
  \hline
  Co. Algmt. & LO\_AI2  \\
  \hline
  Description & Use AI to generate, execute, and adapt pandas snippets for concrete tasks. Students get a CSV file and individually prompt the AI—e.g. P2. They paste the AI’s suggestion in their Jupyter notebook, run it, then manually modify one parameter (e.g. change the plot to a horizontal bar, add labels) to fit a new requirement.   \\
   \hline
  Resources & P2: \emph{“Write pandas code to load a CSV, drop duplicate rows, group sales by region, and plot them. I am going to provide a code template. Everything in ALL\_CAPS is a placeholder. Please preserve the structure exactly.”}  [GENERATE] [Template Pattern]\newline
CSV file with sales data \\
    \hline
  Assessment & Volunteers briefly show their adapted snippet.  \\
  \hline
  \end{tabular}
\\\cline{1-2}

Assignment/Homework/\newline Additional Activities & 
A take-home notebook in which students must:\newline
-- Use an AI assistant to generate pandas code for a specified task. (LO\_AI2)\newline
-- Critically reflect the generated code. (LO\_AI3)\newline
-- Make at least two nontrivial manual adaptations and comment why. (LO\_AI3)
\\ \hline
\multicolumn{2}{|c|}{\cellcolor{04ref} \textbf{4. Lecture Reflection} } \\\hline
  Discussion &
--In what scenarios did the AI help you most? \newline
--Where did you need to intervene or correct its suggestions? \newline
--How will you integrate AI support in your future analyses? \newline
--Can AI tools fully replace manual coding? \\ \cline{1-2}
Critical Thinking &
-- In which situations would I trust AI-generated code without a manual review? \newline
-- How might unnoticed AI-introduced errors impact real business decisions? \newline
-- Could using AI in data analysis introduce bias or privacy risks? How would I detect them?
\\ \hline

\multicolumn{2}{|c|}{\cellcolor{05misc} \textbf{5. Misc. Information} } \\\hline

  References/Material &
Prompt pattern catalog: \url{https://arxiv.org/abs/2302.11382},  
Pandas documentation: \url{https://pandas.pydata.org/docs}   
\\ \hline
\end{tabularx}
\caption{Excerpt of an AI-Blueprint: Pandas Lecture in Python (the full blueprint is available online\footref{fnlabel})}
\label{tab:blueprint}
\vspace{-1em}
\end{table*}

\section{Preliminary Validation}
\label{sec:validation}

As a first step towards validating the applicability and usefulness of our AI-Blueprint, we created an initial set of publicly accessible AI-Blueprints for three lectures within the knowledge areas Software Development Fundamentals, Algorithmic Foundations, and Requirements Engineering. 
Each blueprint was developed by one of the authors, who tailored it to the specific context, LOs, and challenges of a course they teach. Throughout the process, we conducted several internal peer reviews. Feedback was incorporated to improve the structure and clarity of each blueprint. The final blueprints were published as open teaching materials in a repository.

The first blueprint was developed for an \textit{introductory programming course in Python}, focusing on using AI assistants to support students working with pandas. Activities include prompting AI for code explanations, generating data transformation snippets, and evaluating the correctness of AI-generated responses. The second blueprint targets an \textit{algorithms and data structures course}. Students are guided to prompt AI systems to explain algorithmic concepts, compare solutions, and critique AI-generated code. Emphasis is placed on recognising when AI-generated explanations are incorrect or incomplete. The third blueprint integrates AI into an \textit{SE course module on behaviour-driven development}. Activities include generating user stories with AI, comparing AI-generated test cases to manually written ones, and critically assessing coverage and quality. 


Based on these blueprints, we collected feedback and conducted a series of semi-structured interviews with participants from two different universities. The goal was to gain insights and a deeper understanding of the current usage of AI tools and techniques as part of university courses, potential challenges educators face, and whether a structured process and blueprint, together with open-source teaching materials, could alleviate the effort for creating these materials from scratch.

The objective of the preliminary scoping interviews was to gather initial feedback about our proposed process and AI-Blueprints while also identifying opportunities for further improvement and extension. All interview material, questionnaires, and initial version of the blueprints are available online\footref{fnlabel}.

\vspace{0.5em}
\textbf{Interview Setup \& Process:} One researcher created an initial set of questions, which was then discussed among the authors, and the final questionnaire was divided into three parts, with a total of eleven open-ended questions and three Likert-scale questions.
In the first part, we asked participants whether and how they themselves, or their students, used AI tools in their courses. Follow-up questions explored specific AI applications, challenges encountered, and whether any structured approach to AI integration was already in use. After this first part, we briefly introduced the concept of our AI-Blueprints and the blueprint creation process. We deliberately did not introduce our blueprints before part one to avoid any bias when discussing the current state.
The second part focused on presenting the example blueprints and the participants' initial impressions of these blueprints, including perceived usefulness, anticipated strengths and weaknesses compared to their current teaching practices, and willingness to adopt or share such resources. Participants were also invited to comment on missing elements or suggestions for improvement.
Finally, we asked the participants to rate the process, template, and their availability as open materials on a 5-point Likert scale.

To ensure diverse perspectives, we selected participants with varying levels of teaching experience and responsibilities, including early-career PhD students involved in lab sessions and group work, senior lecturers, and professors responsible for entire course modules.
Our study included six participants from two universities involved in software engineering and related courses, including PhD students, senior lecturers, and a professor, with teaching experience ranging from 1 to over 30 years. Their teaching covered a broad spectrum of subjects such as software development, software engineering, software architecture, databases, AI, and algorithms.
Each interview was conducted by at least one researcher in person or via Zoom, lasting approximately 45 minutes. After asking for consent, we recorded all interviews and subsequently transcribed them using the \mbox{OpenAI} Whisper\footnote{\url{https://openai.com/
research/whisper}} speech-to-text service.
After transcribing the interviews, we used open coding~\cite{charmaz2006constructing} where two researchers coded the transcripts, collecting relevant information about (1) current practices and challenges, (2) feedback about the blueprints, and (3) potential improvements and extensions.

\vspace{0.5em}
\textbf{Interview Results:}  The interview participants generally recognised the significant potential of integrating AI tools and activities into courses, particularly within CS curricula. They noted AI's broad usage among students for tasks such as code generation, documentation, and testing, highlighting benefits such as efficiency gains and improved resource access. However, they also expressed concerns about the fact that students are potentially developing an over-reliance on AI tools, which in turn could hamper the development of basic problem-solving skills and critical thinking. Participants emphasised the necessity of clearly communicating the \emph{complementary role of AI}, ensuring that students maintain foundational knowledge and competencies rather than using AI as a complete substitute.

Regarding the proposed blueprint and the structured process, the participants all regarded them positively as valuable resources to enhance teaching effectiveness. Five participants found the process and the template to be helpful (4) or very helpful (5). The openly available teaching material was rated as helpful or very helpful by all interviewees. However, they raised practical concerns about the potential overhead and flexibility of using predefined templates, emphasising the importance of adaptability to individual teaching styles and course requirements. In addition, they suggested the development of a clear didactic framework around AI competencies and task designs, arguing that this would improve the applicability and acceptance of such templates among educators. There was strong support for making these blueprints openly available as teaching resources. However, participants stressed the importance of adaptation and continuous updates aligned with rapid technological developments in the field of AI.

\section{Discussion \& Planned Research}
\label{sec:discussion}

The initial proof-of-concept application of the AI-Blueprints and the responses from the interviews have provided initial evidence that they can be successfully applied to different lectures and course types. Furthermore, the interviews have confirmed that they could be helpful for educators of all experience levels to introduce AI-related topics in their courses without the need to redesign and/or restructure an entire course completely. Although participants acknowledged that the blueprints and process are generally helpful for introducing AI-related tasks, we also received several comments and suggestions for improvements, adding additional details about the LOs, and the structure of how we present individual activities.
Based on the feedback of the interview participants, we identified several areas of opportunity for further research and improvements of our AI-Blueprints.
\vspace{0.5em}

\bullitem{More detailed  Guidelines and Curriculum Integration:}
A participant (PhD level with fewer teaching experience) mentioned that \emph{"the shared blueprints show how the template is meant to be used"} and that \emph{"it might be hard to fill in a new template for a lecture completely from scratch"}. Offering a broader range of AI-Blueprints as open teaching materials for different course categories can support educators in developing their own blueprints by building on existing examples.
However, we are convinced that in addition to providing individual lecture blueprints, in parallel, a broader discussion is required on how to integrate general AI concepts into CS education~\cite{SCHMIDT2025100274,bearman2023discourses}. Although competency models and learning objectives in the education of CS and SE are well established~\cite{malhotra2023shifting}, AI/LLMs as a means to solve SE tasks are still missing. As one concrete future contribution in this direction, we will work towards creating a general competency model for AI-related LOs that can be tailored and adapted to specific course content. Combining this with Bloom's levels to categorize educational goals, or the skill levels described in CS2023, can serve as a starting point for deriving meaningful AI-LOs.\vspace{0.25em}

\vspace{0.5em}
\bullitem{Diversity of AI Activities:}
A second key aspect concerns the diversity of AI-supported activities, which relates to the notion of AI-related LOs discussed above. While LLMs are often associated with content generation, their potential in education should extend beyond that. They can be used for interactive Q\&A, personalised tutoring, feedback generation, and the iterative refinement of exercises, among other applications~\cite{frankford24,cambaz2024use,guha2025future}.

In CS education, activities involving AI and LLMs should extend beyond simple code generation. Particularly in introductory programming courses, students should critically engage with these tools through activities such as reflecting on the reliability and limitations of AI outputs, peer-review supported by model-generated feedback, and assessments that address ethical and integrity considerations. These activities foster technical competence, critical thinking, and responsible AI use.
Aligning such AI-based activities with specific LOs and cognitive levels enables educators to formulate clearer and more measurable AI-related LOs. To support this process, we have started to systematically map common AI activities to relevant cognitive processes to guide the design of aligned learning experiences.

\vspace{0.5em}
\bullitem{Systematic Organisation of Materials:}
Participants found the blueprint template useful for structuring AI-based activities, but highlighted the need for better ways to search and organise them. Particularly for providing open-education materials on a larger scale, better ways of organising, linking them to other existing (non-AI) resources, and presenting the existing blueprints are needed. 
To address this, we are developing a web platform that enables structured exploration of AI-Blueprints, including suitable tools and their alignment with educational tasks. To improve reusability, we plan to categorize activities and blueprints along multiple pedagogical dimensions, such as Bloom’s levels, AI activity type, and skill level.
\vspace{0.5em}

\bullitem{Curricula and Long-Term Perspectives:}
The AI-Blueprints offer a concrete entry point for redesigning lectures, but they also raise broader questions about the continued relevance of existing LOs. As AI increasingly supports routine tasks, curricula must be revisited to determine which competencies remain essential, which should be adapted, and which new objectives should be added~\cite{fernandez2025incorporating}.

 \section{Conclusion}
\label{sec:conclusion}

The emergence of generative AI and LLMs represents both an opportunity and a challenge at the same time, forcing educators to rethink what knowledge and skills students need to acquire in CS and SE courses. Although the rapid growth of these technologies has the potential to improve learning experiences, educators face notable difficulties in coherently embedding AI-related content into existing curricula and courses. Our proposed AI-Blueprints address this need by offering a structured, adaptable framework -- envisioned as open educational resources -- to facilitate the meaningful integration of generative AI technologies into SE lectures and across CS curricula.
Further empirical research is necessary to refine and expand our AI-Blueprints, fostering broader adoption across different educational contexts. 

 
\balance

\bibliographystyle{abbrv}
\bibliography{ICSE2026}

\end{document}